\documentclass[a4paper,12pt]{article}
\usepackage{graphicx}
\usepackage{cite}
\usepackage{amssymb,amsmath,latexsym,enumerate}
\usepackage{subfigure,epsfig}
\usepackage{array,longtable,lscape,bigints}

\begin{document}

\title{On the criteria for integrability of the Li\'{e}nard equation }
\author{Nikolai A. Kudryashov, Dmitry I. Sinelshchikov,\\
\texttt{nakudr@gmail.com}, \texttt{disine@gmail.com}, \vspace{0.5cm}\\
Department of Applied Mathematics, \\ National Research Nuclear University MEPhI, \\ 31 Kashirskoe Shosse, 115409 Moscow, Russian Federation}

\date{}

\maketitle

\begin{abstract}
The Li\'{e}nard equation is of a high importance from both mathematical and  physical points of view. However a question about integrability of this equation has not been completely answered yet. Here we provide a new criterion for integrability of the Li\'{e}nard equation using an approach based on nonlocal transformations. We also obtain some of previously known criteria for integrability of the Li\'{e}nard equation as a straightforward consequences of our approach's application. We illustrate our results by several new examples of integrable Li\'{e}nard equations.
\end{abstract}


\section{Introduction}
In this work we study the Li\'{e}nard equation \cite{Lienard1928,Lienard1928A,Polyanin}
\begin{equation}
y_{zz}+f(y)y_{z}+g(y)=0,
\label{eq:L1}
\end{equation}
where $f(y)$ and $ g(y)$ are arbitrary functions, which do not vanish. Eq.\eqref{eq:L1} is widely used in various applications such as  nonlinear dynamics, physics, biology and chemistry (see, e.g. \cite{Polyanin,Guckenheimer1983,Lakshmanan2003,Andronov} ). For example, some famous nonlinear oscillators such as the van der Pol equation, the Duffing oscillator and the Helmholtz oscillator (see, e.g. \cite{Guckenheimer1983,Andronov,Lakshmanan2005,Frenkel2009}) belong to family of equations \eqref{eq:L1}. What is more, the Li\'{e}nard equation often appears as a traveling wave reduction of nonlinear partial differential equations. Examples include but are not limited to the Fisher equation \cite{Gandarias2013,Kudryashov2014a,Gandarias2015}, the Burgers--Korteweg--de Vries equation \cite{Johnson1970,Kudryashov2009,Polyanin2011} and the Burgers--Huxley equation \cite{Estevez1990,Kudryashov2004,Polyanin2011}. Therefore, it is an important problem to find subclasses of Eq. \eqref{eq:L1} which can be analytically solved.

A problem of the construction of analytical closed--form solutions of Eq. \eqref{eq:L1} has been considered in a few works. For example, integrability of equations of type \eqref{eq:L1} using the Prelle-Singer method was studied in \cite{Lakshmanan2005} and for some particular cases of \eqref{eq:L1} general solutions were obtained.  Lie point symmetries of \eqref{eq:L1} was studied in \cite{Lakshmanan2009,Lakshmanan2009A} and families of equations which can be either linearized by point transformations or integrated by the Lie method were found. Authors of \cite{Mancas2013,Harko2014}  reduced the Li\'{e}nard equation to the Abel equation and used the Chiellini lemma to find a criterion for integrability of Eq. \eqref{eq:L1}. A connection given by non--local transformations between a second--order linear differential equation and equation of type \eqref{eq:L1} was studied in \cite{Meleshko2010}. However, in the above mentioned works not all possible integrable cases of Eq. \eqref{eq:L1} have been found.

In this work we find a new criterion for integrability of the Li\'{e}nard equation. In other words we present a new class of the Li\'{e}nard equations which can be analytically solved. To this end we use an approach that has recently been proposed in \cite{Kudryashov2014,Kudryashov2015,Kudryashov2015a}. Main idea of this approach is to find a connection between studied nonlinear differential equation and some other nonlinear differential equation which has the general closed--form analytical solution. Here we suppose that such a connection is given by means of nonlocal transformations that generalize the Sundman transformations \cite{Meleshko2010,Meleshko2010A,Meleshko2011}. Then we use these transformations in order to convert Eq. \eqref{eq:L1} into a subcase of Eq. \eqref{eq:L1} which general solution is expressed in terms of the Jacobian elliptic functions. We illustrate effectiveness of our approach by providing several new examples of integrable Li\'{e}nard equations. Furthermore, we show that some of previously obtained integrability conditions are consequences of the fact that under these conditions the Li\'{e}nard equation can be linearized by means of nonlocal transformations. To the best of our knowledge our results are new.

The rest of this work is organized as follows. In the next section we present a new criterion for integrability of Eq. \eqref{eq:L1}. We also discuss previously obtained criteria for integrability of Eq. \eqref{eq:L1} in the context of our approach's application. In  Section 3 we illustrate our results by several new examples of integrable Li\'{e}nard equations and construct the general solutions of them. In the last Section we briefly summarize and discuss our results.

\section{Main results}

In this section we consider a connection between Eq. \eqref{eq:L1} and an equation that is subcase of Eq. \eqref{eq:L1} and its general solution can be expressed in terms of the Jacobian elliptic functions. This connection is given by means of the following transformations
\begin{equation}
w=F(y), \quad d\zeta=G(y)dz, \quad F_{y}G\neq0,
\label{eq:L3}
\end{equation}
where $\zeta$ and $w$ are new independent and dependent variables correspondingly.

Among equations \eqref{eq:L1} there is an equation that is of the Painlev\'{e} type and can be solved in terms of the elliptic functions (see, e.g. \cite{Ince}). In this work we study a connection between Eq. \eqref{eq:L1} and this equation that has the form
\begin{equation}
w_{\zeta\zeta}+3w_{\zeta}+w^{3}+2w=0.
\label{eq:L1_3}
\end{equation}
The general solution of \eqref{eq:L1_3} is given by
\begin{equation}
w=e^{-(\zeta+\zeta_{0})}\mbox{cn}\{e^{-(\zeta+\zeta_{0})}-C_{1},1/\sqrt{2}\},
\label{eq:L1_7}
\end{equation}
where $\mbox{cn}$ is the Jacobian elliptic cosine, $\zeta_{0}$ and $C_{1}$ are arbitrary constants. It is worth noting that by scaling transformations Eq. \eqref{eq:L1_3} can be cast into the from $w_{\zeta\zeta}+3\mu w_{\zeta}+\nu w^{3}+2\mu^{2}w=0$, where $\mu$ and $\nu$ are arbitrary nonzero parameters. Since these scaling transformations can be included into \eqref{eq:L3}, without loss of generality we can assume that $\mu=\nu=1$. Note also that there is a singular solution of \eqref{eq:L1_7} which  has the form $w=\pm \sqrt{2}i e^{-\zeta}/(e^{-\zeta}-C_{1})$. Let us finally remark that Eq. \eqref{eq:L1_7} is invariant under the transformation $w\rightarrow -w$ and, therefore, we can use either plus or minus sign in the right--hand side of formula \eqref{eq:L1_7}.

Now we are in position to present our main results.

\textbf{Theorem 1.}  Eq. \eqref{eq:L1} can be transformed into \eqref{eq:L1_3} by means of transformations \eqref{eq:L3} with
\begin{equation}
F(y)=\gamma\left(\int f(y) dy+\delta\right), \quad G(y)=\frac{1}{3}f(y),
\label{eq:L1_5}
\end{equation}
if the following correlation on functions $f(y)$ and $g(y)$ holds
\begin{equation}
g(y)=\frac{f(y)}{9}  \, \left[\int f(y) dy+\delta\right]  \Big( \gamma^{2}\left[\int f(y) dy+\delta\right]^{2}+2 \Big) ,
\label{eq:L1_1}
\end{equation}
where $\gamma\neq0$ and $\delta$ are arbitrary constants.

\textbf{Proof.} One can express $y_{z}$, $y_{zz}$ via $w_{\zeta}$, $w_{\zeta\zeta}$ with the help of \eqref{eq:L3}. Substituting these expressions along with \eqref{eq:L3} into Eq. \eqref{eq:L1} and requiring that the result is \eqref{eq:L1_3} we obtain a system of two ordinary differential equations on functions $F$ and $G$ and an algebraic correlation on functions $f(y)$ and $g(y)$. Solving these equations with respect to $F$, $G$ and $g$ we obtain formulas \eqref{eq:L1_5} and \eqref{eq:L1_1}. Note that one can substitute transformations \eqref{eq:L3} into \eqref{eq:L1_3} and require that the result is \eqref{eq:L1}, which leads to the same formulas \eqref{eq:L1_5} and \eqref{eq:L1_1}. This completes the proof.

One can see that integrability condition \eqref{eq:L1_1} does not coincide with integrability conditions previously obtained in  \cite{Lakshmanan2009,Lakshmanan2009A,Mancas2013,Harko2014}. Using results from works \cite{Lakshmanan2009,Lakshmanan2009A} it can be seen that Eq. \eqref{eq:L1} under condition \eqref{eq:L1_1} admits less than two Lie symmetries, and, thus it can be neither integrated by the Lie method nor linearized by point transformations. Moreover, below we show that condition \eqref{eq:L1_1} is different from a condition for linearizability of Eq. \eqref{eq:L1} via nonlocal transformations. Therefore, condition \eqref{eq:L1_1}  give us a completely new class of integrable Li\'{e}nard equations. In the next section with the help of \eqref{eq:L1_1} we find several new examples of integrable Li\'{e}nard equations.

Now we consider transformations of \eqref{eq:L1} given by means of \eqref{eq:L3} into a linear second order differential equation.

\textbf{Theorem 2.} Eq. \eqref{eq:L1} can be transformed into
\begin{equation}
w_{\zeta\zeta}+\sigma w_{\zeta}+w=0,
\label{eq:L3_3}
\end{equation}
via transformations \eqref{eq:L3} with
\begin{equation}
F(y)=\lambda \left(\int f(y)dy+\kappa \right), \quad G(y)=\frac{1}{\sigma}f(y),
\label{eq:L3_5}
\end{equation}
if the following correlation on functions $f(y)$ and $g(y)$ holds
\begin{equation}
g(y)=\frac{f(y)}{\sigma^{2}}\left[\int fdy +\kappa \right],
\label{eq:L3_1}
\end{equation}
where $\sigma\neq0$, $\lambda \neq0$ and $\kappa$ are arbitrary parameters.

\textbf{Proof.}  One can express $y_{z}$, $y_{zz}$ via $w_{\zeta}$, $w_{\zeta\zeta}$ with the help of \eqref{eq:L3}. Substituting these expressions along with \eqref{eq:L3} into Eq. \eqref{eq:L1} and requiring that the result is \eqref{eq:L3_3} we obtain a system of two ordinary differential equations on functions $F$ and $G$ and an algebraic correlation on functions $f(y)$ and $g(y)$. Solving these equations with respect to $F$, $G$ and $g$ we obtain formulas \eqref{eq:L3_5} and \eqref{eq:L3_1}. Note that one can substitute transformations \eqref{eq:L3} into \eqref{eq:L3_3} and require that the result is \eqref{eq:L1}, which leads to the same formulas \eqref{eq:L3_5} and \eqref{eq:L3_1}. This completes the proof.

One can see that correlation \eqref{eq:L3_1} coincide with one of integrability criteria obtained in \cite{Mancas2013,Harko2014} with the help of the Chiellini lemma.
Therefore, we see that this integrability condition can be found directly from \eqref{eq:L1} without transforming it into the Abel equation. What is more, substituting $f(y)=ay^{q}+b$ or $f(y)=ay+b$, where $a,q \neq 0$ and $b$ are arbitrary parameters, into \eqref{eq:L3_1} we get some of integrability conditions obtained in \cite{Lakshmanan2009,Lakshmanan2009A}. In the case of $f(y)=ay^{q}+b$ we have a subcase of equation \eqref{eq:L1} which admits two Lie symmetries, while in the case of $f(y)=ay+b$ we obtain equation \eqref{eq:L1} with maximal Lie point symmetries. Thus, we see that some of previously known integrability conditions for the Li\'{e}nard equation are consequences of the fact that this equation can be linearized by nonlocal transformations providing that condition \eqref{eq:L3_1} holds. It is also worth noting that correlation \eqref{eq:L3_1} can be found from sufficient conditions for the equation $y_{zz}+\lambda_{2}(z,y) y_{z}^{2}+\lambda_{1}(z,y) y_{z}+\lambda_{0}(z,y)=0$ to be linearizable via nonlocal transformations  \cite{Meleshko2010}.

In this Section a new condition for integrability of the Li\'{e}nard equation has been obtained. It has also been shown that some of previously known conditions for integrability of the Li\'{e}nard equation are consequences of linearizability of the corresponding Li\'{e}nard equation by nonlocal transformations.

\section{Examples}
In this section we provide three new families of integrable Leinard equations. We consider three different cases of the coefficient function $f(y)$: a linear function, a rational function and an exponential function.

\emph{Example 1: a generalized Emden--type equation.}

\begin{figure}[!htp]
\center
\includegraphics[width=0.99\textwidth]{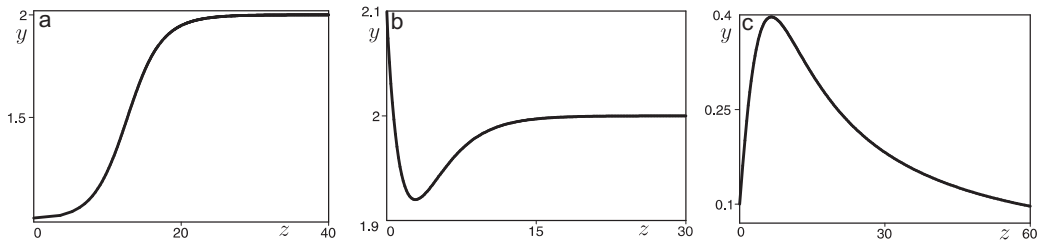}
\caption{Exact solution \eqref{eq:L5_3} of Eq. \eqref{eq:L5_1} corresponding to: a) $\alpha=1$, $\beta=-1$, $y(0)=1.01$, $y_{z}(0)=0$; b) $\alpha=1$, $\beta=-1$, $y(0)=2.1$, $y_{z}(0)=-0.2$; c) $\alpha=1$, $\beta=0$, $y(0)=0.1$, $y_{z}(0)=0.1$.  }
\label{f1}
\end{figure}

Let us suppose that $f(y)=\alpha y+\beta$, $\gamma=2$ and $\delta=0$, where $\alpha\neq0$ and $\beta$ are arbitrary parameters. Then from \eqref{eq:L1_5} we have that
\begin{equation}
\begin{gathered}
F(y)=y(\alpha y+2\beta), \quad G(y)=(\alpha y+\beta)/3,
\label{eq:L5}
\end{gathered}
\end{equation}
and from \eqref{eq:L1}, \eqref{eq:L1_1} we find corresponding Li\'{e}nard equation
\begin{equation}
y_{zz}+(\alpha y+\beta) y_{z}+\frac{y}{18}(\alpha y+\beta)(\alpha y +2\beta)(y^{2}[\alpha y+2\beta]^{2}+2)=0.
\label{eq:L5_1}
\end{equation}
Using \eqref{eq:L3}, \eqref{eq:L1_7} and \eqref{eq:L5} we get the general solution of Eq. \eqref{eq:L5_1}
\begin{equation}
y=\frac{1}{\alpha}\left[\pm \left(\beta^{2}+\alpha e^{-(\zeta-\zeta_{0})}\mbox{cn}\left\{e^{-(\zeta-\zeta_{0})}-C_{1},\frac{1}{\sqrt{2}}\right\}\right)^{1/2}-\beta\right], \,
z=\int \frac{3}{\alpha y+\beta}  d\zeta.
\label{eq:L5_3}
\end{equation}
Let us remark that Eq. \eqref{eq:L5_1} can be considered as a generalization of the modified Emden equation \cite{Lakshmanan2009A} or as a traveling wave reduction of the generalized Burgers--Huxley equation \cite{Estevez1990,Kudryashov2004}.

Plots of solution \eqref{eq:L5_3} corresponding to different initial conditions and at different values of $\alpha$ and $\beta$ are presented in Fig.\ref{f1}. From Fig.\ref{f1} one can see that Eq. \eqref{eq:L5_1} admits kink--type and pulse--type analytical solutions. Note that all solutions presented in Fig.\ref{f1} correspond to the plus sign in formula \eqref{eq:L5_3}.

\emph{Example 2: an equation with rational nonlinearity.}

\begin{figure}[!htp]
\center
\includegraphics[width=0.66\textwidth]{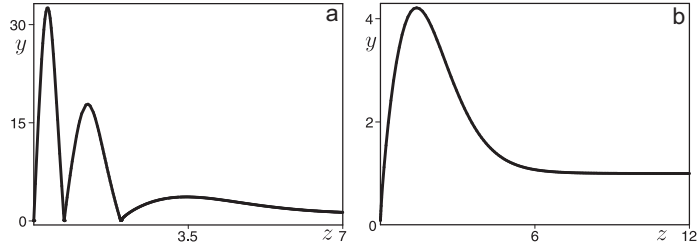}
\caption{Exact solution \eqref{eq:L5_9} of Eq. \eqref{eq:L5_7} corresponding to: a) $\alpha=1$, $\beta=1$, $y(0)=0.025$, $y_{z}(0)=0$; b) $\alpha=1$, $\beta=1$, $y(0)=0.09$, $y_{z}(0)=5$.  }
\label{f2}
\end{figure}

In this example we assume that $f(y)=\alpha/y^{2} +\beta$, $\gamma=1/2$ and $\delta=0$, where $\alpha$ and $\beta$ are arbitrary nonzero parameters. As a result, from \eqref{eq:L1}, \eqref{eq:L1_5} and \eqref{eq:L1_1} we find functions $F$ and $G$
\begin{equation}
\begin{gathered}
F(y)=\frac{1}{2y}(\beta y^{2}-\alpha), \quad G(y)=\frac{\beta y^{2}+\alpha}{3y^{2}},
\label{eq:L5_5}
\end{gathered}
\end{equation}
and corresponding Li\'{e}nard equation
\begin{equation}
y_{zz}+ \left(\frac{\alpha}{y^{2}} +\beta\right)y_{z}+\frac{\beta^{2}y^{4}-\alpha^{2}}{36y^{5}}\left[(\beta y^{2}-\alpha)^{2}+8y^{2}\right]=0.
\label{eq:L5_7}
\end{equation}
The general solution of \eqref{eq:L5_7} is given by
\begin{equation}
\begin{gathered}
y=\frac{1}{\beta}\big(e^{-(\zeta-\zeta_{0})}\mbox{cn}\{e^{-(\zeta-\zeta_{0})}-C_{1},1/\sqrt{2}\}\pm  \hfill \quad \quad \quad \quad \quad \quad \\ \quad \quad \quad\pm\sqrt{e^{-2(\zeta-\zeta_{0})}\mbox{cn}^{2}\{e^{-(\zeta-\zeta_{0})}-C_{1},1/\sqrt{2}\}+\beta\alpha}\big),\\
 z=\int \frac{3y^{2}}{\beta y^{2}+\alpha} d\zeta.
\label{eq:L5_9}
\end{gathered}
\end{equation}
Formula \eqref{eq:L5_9} describes different types of  solutions of Eq. \eqref{eq:L5_7} depending on values of the parameters $\alpha$  and $\beta$ and initial conditions (i.e. values of $\zeta_{0}$ and $C_{1}$). In Fig.\ref{f2} we demonstrate two pulse--type solutions of \eqref{eq:L5_7} which correspond to the plus sign in \eqref{eq:L5_9}.

\emph{Example 3: an equation with exponential nonlinearity.}

Let us assume that $f(y)=\beta e^{\alpha y}$, $\gamma=1$, $\delta=-1$. Then from \eqref{eq:L1_5} we have that
\begin{equation}
F(y)=\frac{\beta}{\alpha}e^{\alpha y}-1, \quad G(y)=\frac{\beta}{3}e^{\alpha y},
\label{eq:L5_11}
\end{equation}
and from \eqref{eq:L1}, \eqref{eq:L1_1} we find corresponding Li\'{e}nard equation
\begin{equation}
y_{zz}+ \beta e^{\alpha y} y_{z}+\frac{\beta}{9\alpha^{3}}e^{\alpha y}(\beta e^{\alpha y}-\alpha)([\beta e^{\alpha y}-\alpha]^{2}+2\alpha^{2})=0.
\label{eq:L5_15}
\end{equation}

\begin{figure}[!htp]
\center
\includegraphics[width=0.66\textwidth]{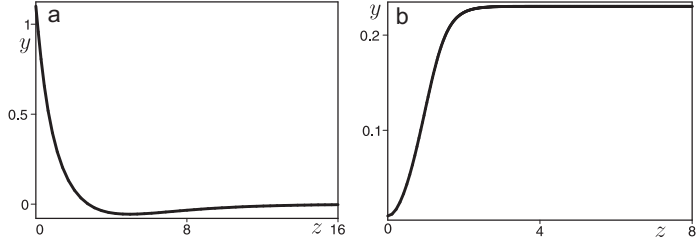}
\caption{Exact solution \eqref{eq:L5_17} of Eq. \eqref{eq:L5_15} corresponding to: a) $\alpha=\beta=1$, $y(0)=1.1$, $y_{z}(0)=-1.1$; b) $\alpha=10$, $\beta=1$, $y(0)=0.01$, $y_{z}(0)=0$.  }
\label{f3}
\end{figure}

The general solution of \eqref{eq:L5_15} can be written as follows
\begin{equation}
\begin{gathered}
y=\frac{1}{\alpha}\ln\left\{\frac{\alpha}{\beta} \left(e^{-(\zeta-\zeta_{0})}\mbox{cn}\left\{e^{-(\zeta-\zeta_{0})}-C_{1},\frac{1}{\sqrt{2}}\right\}+1\right)\right\}, \,
z=\int\ \frac{3}{\beta} e^{-\alpha y} d\zeta
\label{eq:L5_17}
\end{gathered}
\end{equation}
We demonstrate plots of solution \eqref{eq:L5_17} corresponding to different values of the parameters $\alpha$ and $\beta$ and to different initial conditions in Fig.\ref{f3}. It can be seen that Eq. \eqref{eq:L5_15} admits pulse--type and kink--type solutions.

In this section we have presented three new integrable Lienard equations. The general closed--form solutions of these equations have been found. We have demonstrated that these solutions describe various types of dynamical structures.

\section{Conclusion}
In this work we have studied the Li\'{e}nard equation. We have obtained a new integrability condition for this equation. It is worth noting that class of the Li\'{e}nard equations corresponding to this condition can be neither integrated by the Lie method nor linearized by point or nonlocal transformations. We have demonstrated effectiveness of our approach by presenting three new examples of integrable Li\'{e}nard equations. The general solutions of these equations have been constructed and analyzed. We have also shown that some previously obtained integrability conditions follow from linearizabily of the corresponding Li\'{e}nard equations by nonlocal transformations.

\section{Acknowledgments}
This research was supported by Russian Science Foundation grant No. 14--11--00258.

\end{document}